\documentclass{article}
\usepackage{graphicx} 
\usepackage[margin=1in]{geometry}
\usepackage[table]{xcolor}
\usepackage{booktabs}
\usepackage{multirow}
\usepackage{tabularx}
\usepackage{multicol}
\usepackage{enumitem}
\usepackage{sectsty}
\usepackage{hyperref}
\usepackage{times}
\usepackage{titlepic}
\usepackage{float}
\usepackage[multiple]{footmisc}

\usepackage[affil-it]{authblk}
\usepackage[title]{appendix}

\title{\includegraphics[scale=1.5]{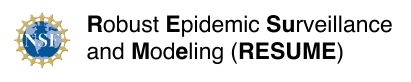}\\[.5cm]NSF RESUME HPC Workshop: High-Performance Computing and Large-Scale Data Management in Service of Epidemiological Modeling}

\author[1,2]{Abby Stevens}
\author[1,2]{Jonathan Ozik}
\author[1,2]{Kyle Chard}
\author[3]{Jaline Gerardin}
\author[1,2]{Justin M. Wozniak}

\affil[1]{University of Chicago}
\affil[2]{Argonne National Laboratory}
\affil[3]{Northwestern University}

\date{}

\begin{document}
\allsectionsfont{\sffamily}

\maketitle
\vspace*{\fill}
\noindent This material is based upon work supported by the National Science Foundation under Grant No. 2200234. We gratefully acknowledge the workshop participants for their engagement throughout the workshop activities and their contributions to this report.

\newpage
\section*{Version Control Table}
\begin{table}[h]
\centering
\begin{tabular}{lll}
\toprule
\textbf{Version} & \textbf{Date} & \textbf{Note}\\
\midrule
v1 & May 25, 2023 & Initial draft \\
v2 & August 3, 2023 & Initial public version \\
\bottomrule
\end{tabular}
\label{table:organizers}
\end{table}

\newpage
\tableofcontents

\newpage
\section*{Executive Summary}
\label{sec:exec_summary}
\addcontentsline{toc}{section}{\nameref{sec:exec_summary}}

The NSF-funded Robust Epidemic Surveillance and Modeling (RESUME) project successfully convened a workshop entitled ``High-performance computing and large-scale data management in service of epidemiological modeling" (HPC workshop for short). This was part of a series of workshops designed to foster sustainable and interdisciplinary co-design for predictive intelligence and pandemic prevention. The event brought together experts in epidemiological modeling, high-performance computing (HPC), HPC workflows, and large-scale data management to develop a shared vision for capabilities needed for computational epidemiology to better support pandemic prevention. 

Through the workshop, we identified key areas in which HPC capabilities could be used to improve epidemiological modeling, particularly in supporting public health decision-making, with an emphasis on HPC workflows, data integration, and HPC access. The workshop explored nascent HPC workflow and large-scale data management approaches currently in use for epidemiological modeling and sought to draw from approaches used in other domains to determine which practices could be best adapted for use in epidemiological modeling. 


The RESUME project is currently transitioning from its planning phase (Phase I) to the potential center phase (Phase II). This workshop contributed to this transition by identifying potential collaborations and laying out work plans to address HPC workflow-related gaps in the predictive intelligence and pandemic prevention landscape. The knowledge and insights gained will be invaluable in shaping future efforts to support and enhance the effectiveness and robustness of pandemic prevention measures.

\subsection*{Summary of the workshop}
\label{sec:workshop_summary}
\addcontentsline{toc}{subsection}{\nameref{sec:workshop_summary}}
The HPC workshop convened 31 invited participants at the University of Chicago on May 1-2, 2023. Participants were drawn from diverse institutions and backgrounds, including epidemiological modelers, software engineers, and high-performance computing (HPC) experts. The workshop spanned two half-days, with the first day focused on identifying key challenges in current computational epidemiology practices and imagining ideal capabilities, and the second day focused on proposing pathways toward developing those capabilities. Programming included a panel and two different configurations of breakout groups, where the bulk of discussions took place. A dinner followed the first workshop session to encourage further discussions.

\subsection*{Summary of discussions and key takeaways}
\label{sec:discussion_summary}
\addcontentsline{toc}{subsection}{\nameref{sec:discussion_summary}}

During the first day of the workshop, participants were shuffled into breakout groups, ensuring each group had participants from both the epidemiological modeling and HPC domains. The modelers in each group were asked to clearly communicate the strengths and weaknesses of their current computational epidemiology approaches to the HPC and large-scale data management experts. Each group then worked together to envision a set of ``ideal world" capabilities. During the next session, participants were asked to select a few of these ideal capabilities and flesh out pathways toward achieving them. Below, we summarize the key discussion points across the primary topic areas and briefly summarize the desired capabilities that were identified and discussed.

\subsubsection*{Strengths, weaknesses, and ideal worlds:}
\begin{itemize}
\item\textbf{Evaluating computational resources for epidemiological modeling:}
Strengths include the use of local and cloud computing, specialized software tools for task management, and substantial storage capacities. User job submission mechanisms and interactive job management features further optimize resource utilization. Weaknesses include data locality issues, portability difficulties among HPC systems, and limitations due to job constraints on scheduled systems. Tedious, manual management of computational processes and the complexity of epidemiological models add to these challenges. For pandemic preparedness in an ideal world, an automated, language-agnostic system would simplify collaboration and enable a science-focused approach. It would provide a comprehensive library of reusable and rated models with clear parameter provenance. This system would have modular problem packaging, facilitate model comparisons, and provide user-friendly interfaces for detailed output. Advanced features would include error-checking automation, performance assessment feedback loops, code containerization, and continuous integration on targeted HPC resources.

\item\textbf{Evaluating current data practices:}
Strengths include effective use and management of diverse datasets, the employment of public data, and well-structured data cleaning practices. Secure enclaves have been established for sensitive data, and automated processes for routine data tasks are in place. Weaknesses include localization problems, lack of automation in some areas, need for more transparent tools, and limitations due to data variability. Legal and security issues also hinder data sharing. Ideally, data storage and access would be secure, seamless, and adaptable. Cleaning and validation would be automated, with data and analysis artifacts, including model outputs and models, being FAIR (Findable, Accessible, Interoperable, Reusable)\footnote{Huerta, E. A., Ben Blaiszik, L. Catherine Brinson, Kristofer E. Bouchard, Daniel Diaz, Caterina Doglioni, Javier M. Duarte, et al. 2023. “FAIR for AI: An Interdisciplinary and International Community Building Perspective.” Scientific Data 10 (1): 487. https://doi.org/10.1038/s41597-023-02298-6.} and easily publishable through research data portals\footnote{Nickolaus Saint, Ryan Chard, Rafael Vescovi, Jim Pruyne, Ben Blaiszik,
Rachana Ananthakrishnan, Michael E. Papka, Kyle Chard, and Ian Foster. 2023. Active Research Data Management with the Django Globus Portal Framework. In Practice and Experience in Advanced Research Computing (PEARC ’23), July 23–27, 2023, Portland, OR, USA. ACM, New York, NY, USA. https://doi.org/10.1145/3569951.3593597}. A standardized data API would ease data retrieval, and clear version control would ensure data reliability. Interaction with HPC resources would be simplified, and the system would prevent the creation of unnecessary intermediate output.

\item\textbf{Evaluating access to technical support:}
Strengths include processes for sanity checks on data ensuring quality and accuracy, and the involvement of scientists in the design of computing systems, tailoring them to specific scientific needs. Weaknesses include a dependency on manual sanity checks, resulting in a lack of automated verification procedures and potential for human error. The lack of long-term computer science personnel due to grant restrictions presents challenges, sometimes causing shortages of necessary tools or training. Ideally, comprehensive technical support would be available, featuring dedicated software engineers with extensive experience, provisioning of tutorials, co-design of computing systems with users, agile short and robust long-term support, and readily available HPC resources. Accomplishing this would involve discussions with HPC facilities personnel, system-level support, GPU support, AI hardware, and ensured data security. Human interaction would be central to the support structure, which would also provide extensive documentation geared towards the different system users.

\end{itemize}

\subsubsection*{Desired capabilities}

\begin{itemize}
    \item \textbf{Automated data retrieval, cleaning, and validation:} An ideal data management system should be seamless and user-friendly, with automated processes taking care of routine tasks such as data retrieval, cleaning, and validation. 
    \item \textbf{Accessible tutorials, trainings, and documentation:} We need clear tutorials and demos targeted to different skill levels for all involved in modeling efforts. This would involve the creation of essential, task-specific documentation in various formats (like text tutorials, videos, and live chats), making the system accessible and user-friendly.
    \item \textbf{Modular and calibrated models for different scenarios with clear provenance for parameters and data sources:} We should have a repository of calibrated models suitable for different scenarios that can be quickly deployed and adapted. Model parameters should have clear provenance.
    \item \textbf{Easy collaboration on commonly used platforms:} We should have language-agnostic, standardized, commonly-used platforms that allow researchers to focus more on science than computation or software. It would make it easy to re-use others' models and their results, fostering collaborative work and work that builds on rather than reinvents others'.
    \item \textbf{Streamlined interaction with HPC resources:} Making interactions with high-performance computing (HPC) resources easier is a key requirement. This could involve features such as low-barrier data staging, alternatives to file systems, streamlined authentication, and visualization capabilities for HPC outputs.
\end{itemize}

\newpage

\section{Introduction}
Large-scale computational epidemiology, supported by high-performance computing (HPC), played a key role in informing decision-making during the COVID-19 pandemic \footnote{Ozik, Jonathan, Justin M Wozniak, Nicholson Collier, Charles M Macal, and Mickaël Binois. 2021. ``A Population Data-Driven Workflow for COVID-19 Modeling and Learning." The International Journal of High Performance Computing Applications 35 (5): 483–99. https://doi.org/10.1177/10943420211035164.}. Epidemiologic models were used to forecast a wide variety of outcomes and test policy interventions for public health stakeholders, often with a very quick turnaround\footnote{Hotton, Anna L., Jonathan Ozik, Chaitanya Kaligotla, Nick Collier, Abby Stevens, Aditya S. Khanna, Margaret M. MacDonell, et al. 2022. ``Impact of Changes in Protective Behaviors and Out-of-Household Activities by Age on COVID-19 Transmission and Hospitalization in Chicago, Illinois." Annals of Epidemiology, June, S1047279722001053. https://doi.org/10.1016/j.annepidem.2022.06.005.}. In addition, there was unprecedented production\footnote{Else, Holly. 2020. How a torrent of COVID science changed research publishing — in seven charts. Nature 588, 7839 (December 2020), 553–553. DOI:https://doi.org/10.1038/d41586-020-03564-y}\footnote{X. Cai, C. V. Fry, and C. S. Wagner. 2021. International collaboration during the COVID-19 crisis: Autumn 2020 developments. Scientometrics 126, 4 (April 2021), 3683–3692. DOI:https://doi.org/10.1007/s11192-021-03873-7} and co-production\footnote{Ray, E.L. et al. 2020. Ensemble Forecasts of Coronavirus Disease 2019 (COVID-19) in the U.S. Epidemiology.}\footnote{Borchering, R.K. et al. 2021. Modeling of Future COVID-19 Cases, Hospitalizations, and Deaths, by Vaccination Rates and Nonpharmaceutical Intervention Scenarios — United States, April–September 2021. MMWR. Morbidity and Mortality Weekly Report. 70, 19 (May 2021), 719–724. DOI:https://doi.org/10.15585/mmwr.mm7019e3.} of scientific work during this time, with researchers across many disciplines eager to help. However, individual research groups generally worked independently as they sought to exploit advances in HPC, data management, machine learning (ML), artificial intelligence (AI), and automation methods when developing, calibrating, modifying, verifying, and validating epidemiologic models. 

The NSF-funded Robust Epidemic Surveillance and Modeling (RESUME) project seeks to, among other things, lower the barriers to and automate epidemiologic model analyses, monitoring, and rapid response on HPC resources. To support this effort, we hosted the workshop ``High-performance computing and large-scale data management in service of epidemiological modeling," which brought together experts in modeling, HPC, HPC workflows, and large-scale data management to develop a shared vision for capabilities needed for computational epidemiology to better support pandemic prevention. This two-day workshop provided an opportunity for cross-disciplinary experts to share their perspectives, needs, and constraints with one another, and outline their ideal-world capabilities. We identified key areas in which HPC capabilities could be used to improve epidemiological modeling, particularly in supporting public health decision-making, with an emphasis on HPC workflows, data integration, and HPC access. The workshop explored nascent HPC workflow and large-scale data management approaches currently in use for epidemiological modeling and sought to draw from approaches used in other domains to determine which practices could be best adapted for use in epidemiological modeling. In addition, through the workshop, the RESUME project sought feedback on the design and development of the Open Science Platform for Robust Epidemic analYsis (OSPREY), which the RESUME team is actively developing. OSPREY is being built to leverage investments in forthcoming exascale and increasingly ubiquitous HPC and data resources\footnote{Collier, Nicholson, Justin M. Wozniak, Abby Stevens, Yadu Babuji, Mickaël Binois, Arindam Fadikar, Alexandra Würth, Kyle Chard, and Jonathan Ozik. 2023. ``Developing Distributed High-performance Computing Capabilities of an Open Science Platform for Robust Epidemic Analysis" 2023 IEEE International Parallel and Distributed Processing Symposium Workshops (IPDPSW), St. Petersburg, FL, USA, 2023, pp. 868-877, doi: 10.1109/IPDPSW59300.2023.00143.}.

As the RESUME project moves from its current planning phase (Phase I) to a potential center phase (Phase II), this and other workshops will provide an opportunity to identify collaborations and more effectively address gaps in the current predictive intelligence and pandemic prevention landscape.

\section{Workshop structure and activities}
The HPC workshop brought together 31 researchers from a wide variety of institutions and research areas for a \mbox{2-day} event that took place at the University of Chicago. For a complete list of participants and their affiliations, see Appendix \ref{tbl:participants}. The workshop programming and logistics were developed by a 5-member organizing committee (Table~\ref{table:organizers}), combining both RESUME project facilitators and subject matter experts.

\begin{table}[h]
\centering
\rowcolors{2}{gray!25}{white}
\begin{tabular}{lp{3cm}p{7cm}}
\toprule
\textbf{Role} & \textbf{Name} & \textbf{Affiliation}\\
\midrule
Subject matter experts & Kyle Chard \newline Jaline Gerardin \newline Justin Wozniak & 
    University of Chicago, Argonne National Lab \newline Northwestern University \newline University of Chicago, Argonne National Lab\\
RESUME project facilitators & Jonathan Ozik \newline Abby Stevens & University of Chicago, Argonne National Lab \newline University of Chicago, Argonne National Lab \\
\bottomrule
\end{tabular}
\caption{Workshop organizers}
\label{table:organizers}
\end{table}

The goal of the HPC workshop was to work together to develop a shared vision for the ways in which HPC capabilities can be leveraged to better support pandemic prevention. Specifically, we aimed to understand challenges and desired capabilities for computational epidemiology workflows. Programming consisted of a panel and two sessions of breakout groups, each of which concluded with short presentations to the broader workshop. The complete agenda can be found in Appendix \ref{tbl:agenda}.

\subsection{Workshop Participant Panel}
\begin{figure}[t]
\centering
\includegraphics[width=\textwidth]{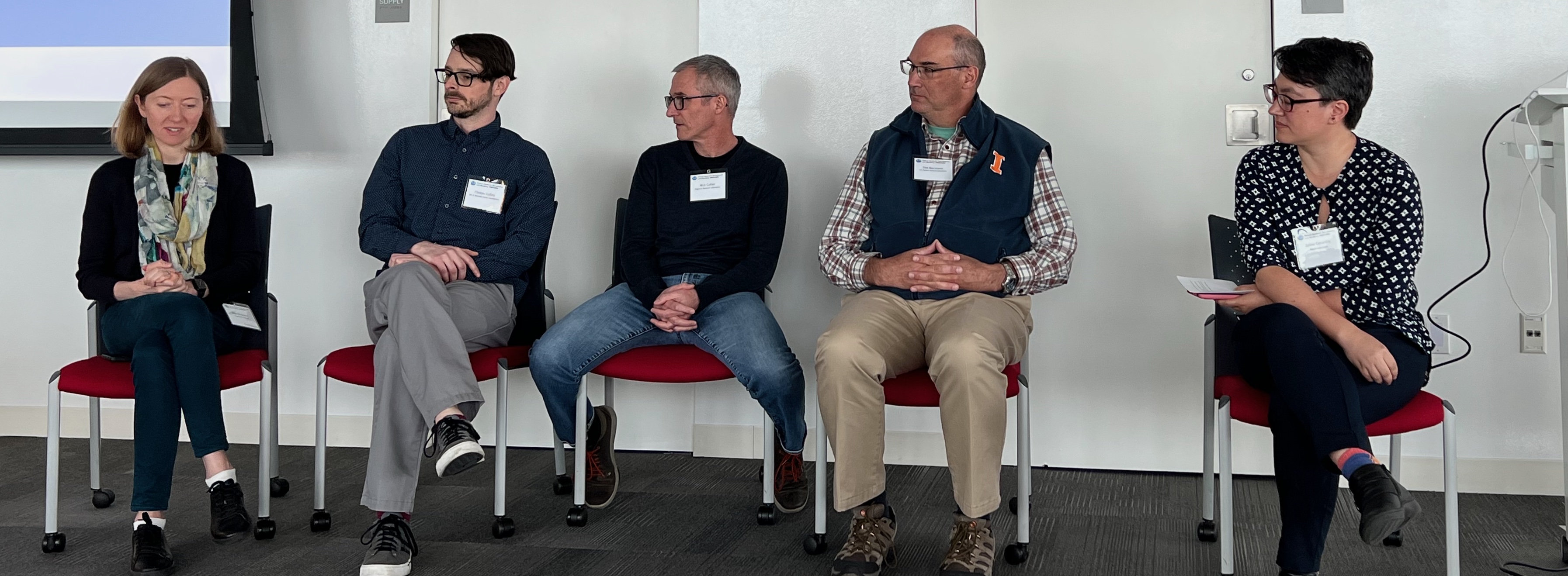}
\caption{Panelists Olga Morozova, Clinton Collins, Nick Collier, and Tim Germann with moderator Jaline Gerardin (left to right)}
\label{fig:panel}
\end{figure}

A panel of experts representing various sides of computational epidemiology teed off the core content of the workshop with a discussion and evaluation of their current computational epidemiology practices (Figure \ref{fig:panel}). Our panelists provided valuable insights into key challenges and desired capabilities across their disciplines to motivate and frame this workshop.

\subsubsection*{Panel structure}
\begin{itemize}
    \item[] \textbf{\textit{Moderator}}
    \begin{itemize}
        \item[] \textbf{Jaline Gerardin}, Assistant Professor of Preventative Medicine, Northwestern University
    \end{itemize}
    \item[] \textbf{\textit{Panelists}}
    \begin{itemize}
        \item[] \textbf{Olya Morozova}, Assistant Professor, Public Health Sciences Department, University of Chicago
        \item[]\textbf{Clinton Collins}, Software Engineer Manager, Bill \& Melinda Gates Foundation
        \item[] \textbf{Nick Collier}, Senior Software Engineer, Argonne National Laboratory/University of Chicago
        \item[] \textbf{Tim Germann}, Technical Staff Member, T-1, Los Alamos National Laboratory
    \end{itemize}
\end{itemize}

\subsubsection*{Key takeaways}

The panelists had all been involved in designing and deploying computational epidemiology systems in response to the COVID-19 pandemic. While many takeaways from the panel emphasize possibilities for future improvements, it is worth noting that the systems they designed under the extreme pressure of the early pandemic were all able to be used by public health officials for pandemic response and decision-making. Panelists agreed that this was a noteworthy accomplishment, given the constraints. That said, the following topics were highlighted as areas of significant interest and with room for improvement before the next pandemic.

\begin{itemize}

\item \textbf{Data management:} A key challenge emphasized by panelists was the availability, harmonization, and standardization of the data sources required for COVID-19 modeling. There were known and unknown biases in data, particularly around things like case counts and hospitalizations. A lot of data was either inaccessible or not available in a format amenable to automation, which led to teams spending valuable time creating data scrapers. Furthermore, complications arose in data sharing due to legal and organizational restrictions, resulting in teams engaging with legal groups more than anticipated and leading to long waits for data before modeling could begin. Frequent modifications to data formats, such as hospitals renaming columns in their spreadsheets week-to-week, made it difficult to fully automate pipelines. Panelists agreed that setting community standards around data would have significantly streamlined modeling efforts.

\item \textbf{Model development:} A variety of model types were deployed by panelists and their teams. Some were starting from scratch, while others were able to leverage existing tools. Based on the fluid and changing nature of the pandemic, models needed to be flexible and quickly adapted, often requiring re-calibrations that could be computationally or time-intensive. Some panelists felt that their models were ultimately unnecessarily complex, which was not something they were able to anticipate at the start of development. All agreed that some kind of model repository or open modeling framework would help streamline future model development and unify efforts.

\item \textbf{Software and workflows:} Computational epidemiology systems rely heavily on workflows to be robust and software to be efficient at the scale of large computing systems. However, many researchers lack sufficient support for building out complete systems, requiring them to spend significant time learning to run 3rd-party software or manually operate computational systems. Researchers that did have access to HPC systems often faced issues with model deployment and debugging. They highlighted the lack of standardization among workflow systems, pointing out that there are numerous different systems in use, although a standardization attempt could result in the creation of even more systems. In an ideal world, they would develop a workflow that is compatible with different architectures in various labs, universities, and government institutions. They also suggested the need for automated processes in these workflow systems. An ideal automated system could streamline data updates, adjust model parameters, and launch simulations, thus saving time and enhancing efficiency.

\item \textbf{Collaboration:} Panelists emphasized the importance of better collaboration with stakeholders, along with the establishment of more clearly defined roles within modeling teams, including more HPC experts and designated software engineers. They also touched on challenges of collaboration in academia, especially in the context of correctness-critical, fast-paced contexts like a pandemic. The academic culture of innovation and building things from scratch made it such that using others' software could be seen as less innovative and could potentially affect career progression and prestige. From a public health perspective, panelists agreed that in the long term, the operational support for these modeling efforts should be provided by public health departments rather than academia.
\end{itemize}

\subsection{HPC Breakout Sessions}
The core output of this workshop is a set of desired capabilities to improve the use of HPC systems in computational epidemiology, along with proposed pathways toward achieving these capabilities with the potential to be addressed by the RESUME project. To identify these capabilities and solutions, the workshop facilitated two break out sessions:

\begin{itemize}
    \item \textbf{Session 1: Identifying desired capabilities:} The goal of the first breakout session was for epidemiological modelers to clearly articulate their current practices and challenges to HPC and large-scale data management experts. The session consisted of an evaluation phase, a visioning phase, and a drawing phase. During the evaluation phase, modelers were asked to describe and evaluate their current computational resources, data practices, and technical support, using one modeling project as a representative example. Then, during the visioning phase, they were asked to imagine the capabilities they might have in an ``ideal world." Finally, participants were asked to fully describe and draw their ideal computational epidemiological systems (Figure \ref{fig:drawings}). Following this session, workshop organizers synthesized findings across groups to identify a set of overlapping desired capabilities to further explore. 

\begin{figure}[t]
\centering
\includegraphics[width=\textwidth]{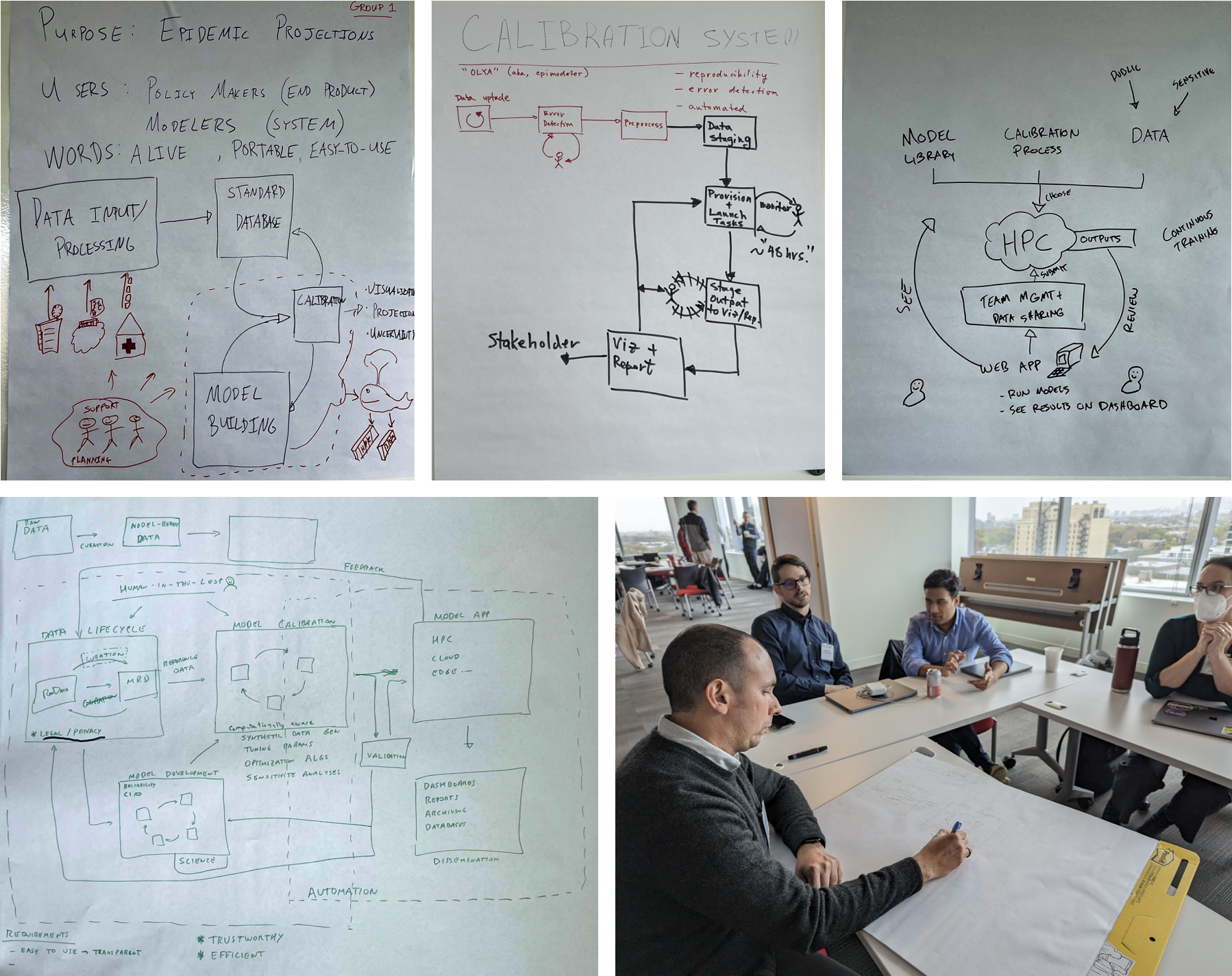}
\caption{Drawings of ``ideal systems" envisioned by the breakout groups, and a glimpse into the process.}
\label{fig:drawings}
\end{figure}

\item \textbf{Session 2: Proposing solutions:} The second breakout session was focused on proposing pathways toward achieving the desired capabilities identified in the first session. Groups were asked to choose between two and four of the synthesized topics to discuss the research developments, data, and technologies needed to bridge the gap between current practices and ideal-world capabilities. Then, participants were asked to write a brief summary of the discussion surrounding each topic, which are included later in this report.
\end{itemize}

\noindent Each breakout group was assigned a RESUME project-affiliated facilitator and notetaker to document the session, and one HPC-focused participant per group was tasked with presenting a summary of their discussion to the broader workshop. A complete list of all questions asked during this session can be found in Appendix \ref{breakout1_qs}. Below, we summarize the key topics and findings from these sessions.

\subsubsection*{Summary of Breakout Session 1}
    \begin{enumerate}
        \item \textbf{Evaluating current computational resources for epidemiological modeling}
        \begin{itemize}
            \item \textbf{\textit{Strengths}:} Many researchers have access to some kind of large-scale computing resource, and most successfully leverage both local and cloud computing resources for efficient computation. The systems often come with sizable storage capacities, capable of handling the vast amounts of data typically involved in large-scale epidemiological modeling. Several groups have developed specialized software tools to interact with these systems.
            \item \textbf{\textit{Weaknesses}:} Data locality issues hamper efficient data analysis in some setups, while others face limitations due to job constraints or the complexity of using multiple versions of job scheduling tools with different plugins. Transitioning from one HPC system to another is often fraught with difficulties, primarily due to understanding differences in performance, managing technical debt, and a lack of proper documentation and training. Many users reported struggling with efficiently using resources and tuning their systems for optimal performance. Furthermore, areas that could benefit from automation, like hyperparameter checking and model calibration, are often manually handled, adding to the time and complexity of the modeling process.
            \item \textbf{\textit{Ideal world capabilities}:} In an ideal world, a language-agnostic, highly automated system would allow for easy collaboration and a focus on science over computation. This system would feature a comprehensive model library with associated code, documentation, and parameter repositories. It would promote model reuse, offering a catalog of pre-rated models for diverse requirements. Essential features would include modular problem packaging, a range of calibrated models for rapid deployment, and clear parameter provenance. This system would also facilitate model comparisons, integrate HPC-friendly calibration capabilities, and provide shared result visualization tools. Automations would handle error-checking, task termination, and retries, with feedback loops for performance assessment. A high level of user-friendliness would allow easy extraction of detailed modeling output, alongside the ability to switch between different model formulations. Other attributes would include the containerization of code, interactive cloud systems, continuous integration on targeted HPC resources, and a software abstraction layer for flexibility and extensive testing.
        \end{itemize}
        \item \textbf{Evaluating current data practices}
        \begin{itemize}
            \item \textbf{\textit{Strengths}:} Many current practices demonstrate effective use and management of diverse datasets. Public data is frequently used, and there are some well-structured practices to clean and prepare these data for further processing. For sensitive data, secure enclaves have been developed to store and analyze data while preserving privacy and confidentiality. Other data practices involve establishing collaborative agreements to access proprietary data. Automated processes are in place to handle routine data tasks, such as downloading. Access to HPC resources and dedicated queues have been identified as strengths. There are also promising efforts to develop better data tools, such as systems for real-time data summary and quality checking.
            \item \textbf{\textit{Weaknesses}:} Data localization is a major problem, as data is often too large or sensitive to move, which may hamper accessibility and computation. There's also a lack of automation in some parts of the data pipeline, which increases manual labor and the potential for error. Several participants reported a need for more transparent and user-friendly tools and straightforward ways to explore the data with model development in mind. Issues related to tracking data and model parameter provenance, data cleaning, and interpretation are common. Data sharing is also an issue, with privacy issues and academic competition often a limitation.
            \item \textbf{\textit{Ideal world capabilities}:} In an ideal world, data storage and access would be secure, seamless, and require minimal user attention, while adapting to specific requirements, such as protected health information. Intentional databases would streamline data management, with cleaning, validation, and quality control processes automated for efficiency. Model outputs would be standardized, tracked, and openly accessible in a FAIR (Findable, Accessible, Interoperable, Reusable) manner, allowing for data reuse by different stakeholders. A standardized data API would facilitate the harmonizing of data from various sources, and automated data retrieval would expedite access. Clear version control and provenance would ensure data reliability. The system would enable the availability of data where needed, either stored within databases or as flat files. APIs would manage access to protected data based on user permissions. Interaction with HPC resources would be made easier, with low-barrier data staging, alternatives to file systems, streamlined authentication, and visualization capabilities for HPC outputs. The system would be designed to prevent the creation of unnecessary intermediate output.
        \end{itemize}
        \item \textbf{Evaluating access to technical support}
        \begin{itemize}
            \item \textbf{\textit{Strengths}:} Several research groups have access to dedicated, long-term software engineers with expertise in both computing and its epidemiological applications. Close collaboration between scientists and computing experts ensures that the systems are designed with an understanding of the specific requirements and constraints of the scientific problems they are intended to solve. When resources are available for such personnel, these collaborations are commonly a huge success.
            \item \textbf{\textit{Weaknesses}:} In many cases, a significant challenge is the lack of long-term computer science personnel, such as staff software engineers and HPC experts with extensive experience, due to short-term funding. This could limit the development and maintenance of sophisticated tools and systems needed for large-scale data handling and modeling. Without dedicated computer science professionals, there might be a shortage of necessary tools for training or a lack of access to those that already exist.
            \item \textbf{\textit{Ideal world capabilities}:} In an ideal world, comprehensive technical support would include access to dedicated staff software engineers, with extensive experience in the models used and deep institutional and domain knowledge, able to implement solutions with minimal instruction. Support services would range from the provisioning of tutorials, demos, and training at all levels, to the co-design of compute systems with end users from the start. Agile, short-term responses would complement robust long-term support, underpinned by support staff who can effectively listen to and communicate with modelers. Readily available HPC resources, reservations, and allocations would be essential, as well as dashboard-type interfaces for managing and investigating model runs. The planning stages would involve discussing the whole workflow with HPC personnel, and there would be system-level support, including GPU support and AI hardware to support novel  calibration and emulation tasks. Security for protected data, possibly through an API, would be ensured. Human interaction would remain central to the support structure, which would also provide essential generic and task-specific documentation in various formats, including text tutorials, videos, and live chats.
        \end{itemize}
    \end{enumerate}

\subsubsection*{Desired capabilities}
From these discussions, workshop organizers synthesized key takeaways to identify desired ideal-world capabilities across the breakout groups (Table \ref{tbl:capable}). They were organized according to scope - some were realistic to address as part of the pilot projects for the current planning phase of the RESUME grant, some were within the scope of the Phase II center grant, and some were long-term goals. 

\begin{table}[h]
\begin{tabularx}{\textwidth}{|>{\centering\arraybackslash}X|>{\centering\arraybackslash}X|>{\centering\arraybackslash}X|}
\hline
\textbf{\textit{Within scope of RESUME pilots}} & \textbf{\textit{Within scope of RESUME center}} & \textbf{\textit{Beyond RESUME scope}} 
\\
\hline
\begin{flushleft}
\begin{itemize}[leftmargin=10pt]
\item Automated generation of reports and visualizations
\item Language-agnostic, flexible modeling systems
\item HPC-friendly calibration frameworks and plug-ins
\item Automated data retrieval, cleaning, and validation
\item Streamlined interaction with HPC resources
\item Accessible tutorials, trainings, and documentation
\end{itemize} 
\end{flushleft} &
\begin{flushleft}
\begin{itemize}[leftmargin=10pt]
\item Easy collaboration on commonly-used platforms
\item Repository of model parameters for reference
\item Facilitated reuse of models and results
\item Modular and calibrated models for different scenarios
\item Clear provenance for parameters and data sources
\item Interactive cloud systems with continuous integration
\item Streamlined data staging for modeling, analysis, and visualization
\end{itemize} 
\end{flushleft} &
\begin{flushleft}
\begin{itemize}[leftmargin=10pt]
\item Standardized data APIs among sources
\item Closer collaboration with surveillance networks
\item Comprehensive library of models with code and documentation
\item Experienced technical support and long-term staff
\end{itemize}
\end{flushleft} \\
\hline
\end{tabularx}
\caption{Desired ideal-world capabilities identified across breakout groups during first breakout sessions. Workshop organizers synthesized discussions and sorted capabilities based on scope within the RESUME project.}
\label{tbl:capable}
\end{table}

\subsection*{Summary of Breakout Session 2}
On the second day of the workshop, participants were shuffled into new breakout groups with the goal of proposing pathways toward two-to-four desired capabilities in Table \ref{tbl:capable}. Members of each group then spent time writing up their solutions.  Below, we summarize these write-ups, grouped by desired capability. 


\subsubsection*{\textit{Accessible tutorials, trainings, and documentation / Streamlined interaction with HPC resources}}

Generating accessible tutorials, trainings, and documentation is essential for maximizing the impact of HPC-based epidemic modeling. While there are general skills and techniques that might be useful to a wide audience, we concluded that linking training to a concrete product, such as a plug-and-play model calibration framework, would be the best use of time and resources. 

We identified two primary audiences for trainings and tutorials related to HPC applications for epidemic modeling. First, we imagined tutorials that would help modelers make the best use of the new tools and frameworks that we develop. For instance, one could imagine tutorials that illustrate how a modeler might code an objective function for their specific problem in a way that maximizes compatibility with different calibration methods. Second, we imagined that educational materials could be developed to facilitate conversations between modelers and other stakeholders, such as policy makers. As a concrete example, interactive demonstrations could be developed to allow modelers to quickly illustrate different models and calibration methods to a policy maker. We likened this to a kind of A/B testing which would allow end-users to make informed decisions about modeling approaches given the specific deliverables that they need. One exciting future direction that was discussed was using AI tools to aid end users in understanding complex code bases. As large language models such as ChatGPT become more powerful, these could become invaluable tools that would allow an end user to query the codebase for a complex epidemic model in order to understand what it does. 

Tutorials and trainings should go beyond simple text-based descriptions and aim for interactivity. For example, using interactive notebooks as a way to integrate code and example outputs could be a powerful way to keep documentation fresh and useful for end-users. The following is a proposed vision for a centralized help resource that would provide general and discipline-specific guidance for introductory, intermediate, and advanced users of HPC resources within and across institutions. The resource (which can be considered as a type of wiki or website) would include the following components:

\begin{itemize}
\item \textbf{Interactive Resources for First-time Users:} First-time users who have limited familiarity with the command line, computing clusters, or version control, would be directed to tutorials that use interactive R and Python coding (such as RMarkdown and Jupyter notebooks) that can be run directly on the cluster.
\item \textbf{Links to Introductory Resources for Advanced Beginners:} For beginning users who will eventually be writing their own submission scripts and can devote more time to training, the website will recommend a series of introductory trainings in Unix, GitHub, and basic HPC job submission (leveraging existing resources such as Data Carpentry). These trainings could also be conducted in person when possible, perhaps initially by HPC staff and then later by experienced HPC users, ideally integrated with other data science training initiatives. 
\item \textbf{Resources for Intermediate Users:} Modelers often need to use particular types of parallelization for their code. A major challenge that more advanced users encounter when debugging is that solutions that can be found on web platforms (such as Google or Stack Overflow) may be specific for a cluster configuration that is different from the cluster they are using (for example, SLURM vs. other configs). One solution to this challenge would be to create a list or table  of common issues encountered by users, with solutions for each type of cluster. This information could be compiled by HPC staff across or within institutions and clusters. 
\item \textbf{Project Archive for Advanced Users:} Advanced users may require particular discipline-specific modifications to their workflows. Having a curated archive of past implemented discipline-specific projects could be very helpful in this regard. Each project could contain a description of the overall motivation, scripts and code used, any packages, dependencies, and environments (or perhaps containers), as well as the name of the PI and the HPC staff member involved in shaping the project. The archive could link to publications and GitHub repositories and contain tags for particular disciplines or types of jobs/software/workflows, and would provide investigators with HPC staff with domain expertise whom they could contact while showcasing the breadth of projects enabled by the HPC. 
\item \textbf{FAQ section for troubleshooting:} Members of participating institutions with further questions could submit their query in a model similar to a Stack Overflow post. However, they would also tag or reference their discipline, cluster type, and  relevance to past examples in their query. The answers to these questions would create another resource for advanced users who encounter issues.

\item \textbf{Project Feedback Form:} PIs who are members of participating institutions could submit a feedback form to their respective HPC facility containing information about their proposed project including the aims, scope, and details on the type of software, packages, and type of parallelization that might be required. This form would be reviewed by an HPC staff member with domain expertise who could provide quick feedback on the feasibility of the project and highlight potential problems that may arise with packages or dependencies. They might even help develop a container for the PI, a preliminary sandbox on the HPC resource that the PI can use for development, or a virtual environment that emulates the HPC resource that the PI can put on their local machine. 
\end{itemize}

Once more specific HPC-based modeling tools are developed, there was enthusiasm about the idea of creating summer programs to help educate potential end users on how to use them. Ideally, if the tools developed adequately abstract away the platform-specific aspects of HPC, these trainings could focus on the general principles of HPC and go more in-depth on building and comparing different epidemic models at large scales. It was recognized that there are many existing educational materials related to epidemic modeling in an HPC environment. Research and information-gathering would need to be done to collect relevant courses on platforms such as Udemy and Coursera that could serve as valuable references for someone new to the field of HPC computing or epidemic modeling. These resources could be collected on a website for anyone to easily access. 

\subsubsection*{\textit{Automated data retrieval, cleaning, and validation}}

The COVID-19 pandemic has highlighted the need for broad, accurate, and timely data in informing public health decisions. However, the data used to inform epidemiological models is often sourced from many disparate sources, making it difficult to ensure that data are accurate, consistent, and up-to-date. Significant time is spent by modelers to identify, acquire, align, curate, and validate data and, thus, it is an area ripe for automation. An automated platform would streamline the data collection process, reducing the time and resources required to gather data from various sources. Further, such a platform would automate the process of cleaning and validating the data, ensuring that it is accurate and consistent across all sources. The platform could automatically check the quality of the data, flagging any inconsistencies or errors that need to be addressed. In such cases, it would engage humans to review data quickly (e.g., through visualizations of data distributions) and identify inconsistencies or errors prior to use. The platform would also automate the process of merging multi-modal data from multiple sources and formats, aligning to a common representation for model input---which is often a time-consuming and error-prone task. This would ensure that data is consistently updated and readily available for use in models. 

 The following is a vision for a \textbf{Bio-preparedness Data Hub}, which would automate data retrieval, cleaning, and validation that is crucial for improving the accuracy and reliability of COVID-19 and other infectious disease models. This would ultimately lead to better public health decisions and improved outcomes for individuals and communities impacted by the pandemic. Implementing the DataHub requires addressing the following key areas.  

\begin{itemize}
\item \textbf{Extensible Connectors to Data Sources:} A system must be implemented to connect to arbitrary external data sources such as public health agencies, laboratories, and other relevant institutions. These data may be distributed in different formats, with different protocols. Thus, the system must support common connector types. There would be the need for users to be able to define schema mappings from the remote format to their desired common format. Finally, it is crucial that metadata be collected and stored along with the data (e.g., data acquisition time, source, checksum).
 
\item \textbf{Automate Review/Download:} Data are released sporadically. The DataHub must be able to automatically detect when data are released or updated and then start a process to reliably download the data into the target environment. If any unexpected changes to the file format are detected, users will be notified so any code can be appropriately updated. As part of this process the data would proceed through an ingestion pipeline to visualize, compare, and integrate into a common schema. 
 
\item \textbf{Visualization with Humans in the Loop:} Visualizing data is an essential part of understanding it. The DataHub would automatically generate visualizations such as distributions and figures providing a high-level, aggregated, and quickly digestible view of the data. Such approaches would allow modelers to quickly identify changes and anomalies in the data. This crucial step in the pipeline would facilitate human-in-the-loop validation for critical data. 
 
\item \textbf{Data Source Comparison:} There are likely to be overlapping data from different sources, these sources may have varying degrees of accuracy, granularity, and freshness. The DataHub would automatically compare overlapping data from different sources and identify any inconsistencies or discrepancies. It would engage humans to select or prioritize data sources for conflicts and notify users of inconsistencies.  
 
\item \textbf{Data Integration:} Target data are likely to be multi-modal, heterogeneous, and delivered with diverse schema. To consume these data in models they must first be joined, aligned, and then cleaned in user-defined ways. As part of this step, data with different spatial and temporal resolutions could be aligned to facilitate analysis. Users would be able to define how these data are to be integrated. Importantly, data change over time. The DataHub would need to store both provenance capturing when, where, and how data are derived and also store version history such that users can interrogate the provenance of data and also facilitate reproducibility (e.g., to document what data were used in a paper or presentation).  
 
\item \textbf{Continuous Integration:} Finally, the DataHub would continuously validate the data based on user-defined rules. This could include checking for outliers, identifying missing data, and ensuring that the data meets established standards. The continuous integration process would be triggered automatically based on changes to the data or models, periodically, or by users. 
\end{itemize}

\subsubsection*{\textit{Modular and calibrated models for different scenarios}}

A central repository of models with calibrated parameters to be used out-of-the-box for a variety of epidemiological scenarios would be an immense time-saver in times of crisis. However, it was noted that criteria would need to be developed for adding models to the repository, such as whether they have passed peer-review, provide source code and document dependencies, include datasets and validation cases, and facilitate reproducibility. 
 
The purpose of such a repository could be to not only be a source for models, but also to facilitate the sharing and discovery of models. It could facilitate the establishing and improving of standards practices, capabilities of scale, expertise in calibration and validation, for example.  The repository would not only be for the models, but also a place where users of the model describe how they have used the models, and a repository for use cases. The repository could be used to provide opportunities for using, advancing, and teaching about the models.


\subsubsection*{\textit{Clear provenance for parameters and data sources}}
A calibrated model is only useful if its provenance is known - this includes model parameters, calibration data, software use and runtime environment information. It is impractical to collect this information manually - instead, provenance should be automatically captured during the data gathering, calibration, software development and model execution. Doing this enables models to be quickly and easily understood and reused.

The data used for calibration should have proper provenance attached including its source, any preprocessing steps taken to clean the data, and some kind of validation. When actually calibrating the model, there should be detailed documentation on the tools and methods used, in addition to the result. To capture the software/runtime environment, the code repository should include relevant details for full reproducibility. Clear, well-documented models and model parameters would provide an invaluable resource for computational epidemiology.

\subsubsection*{\textit{HPC-friendly calibration frameworks}}

In the context of HPC-friendly calibration frameworks, several technology-related items were identified as addressable goals. One important aspect is to make existing algorithms more accessible and comparable, with the ability to easily integrate different models and objective functions. Calibration frameworks must also be resource-aware, taking into account the scheduling of resources on compute clusters, supporting asynchronous workflows, and different job execution modes (batch, iterative, etc.). Pilot-job based systems were identified as having the potential to better support the variety of calibration approaches, particularly online methods. Related to calibration, documentation on the best practices such as common calibration techniques, workflows, and algorithms would be useful. This could include a comprehensive list of calibration methods, including their advantages, disadvantages, and the specific problems they are best suited for.

\section{Key takeways}
Computational epidemiology, facilitated by HPC systems, has the potential to be an essential tool in future pandemic prevention. Throughout this workshop, participants shared and discussed the extensive work they did in response to the COVID-19 pandemic, all done under high-pressure constraints to support urgent public health decision-making. With time to reflect, it was clear that researchers around the world were responding to similar requests that could be addressed using common tools, but few of those existed and there was little time to build them during the height of the pandemic, when they would have been most useful.

When asked to envision their ideal computational systems for epidemiology, many common themes emerged across workshop discussions. Ideal computational systems would be user-friendly, collaborative, and as automated as possible. Epidemiological models need not be developed from scratch for each specific use case; many elements can be developed ahead of time for easy reuse. Models that have been successfully deployed should be thoroughly documented and included in accessible repositories. HPC systems should be easily accessible and available to researchers; it should not be a requirement that you need to be an HPC expert to leverage HPC systems for computational epidemiology.

As the RESUME project moves beyond the planning phase, these insights are helping to inform future developments. In particular, the RESUME team is currently developing the Open Science Platform for Robust Epidemic analYsis (OSPREY) and the discussions from this workshop are being used to inform the use-inspired development of its capabilities. 

\newpage
\begin{appendices}
\section{Appendices}

\subsection{Workshop Participants}\label{participants}

\begin{table}[h]\label{tbl:participants}
\centering
\rowcolors{2}{gray!25}{white}
\begin{tabular}{ll}
\toprule
\textbf{Name} & \textbf{Institution} \\
\midrule
Abby Stevens                 & University of Chicago/Argonne National Laboratory                          \\
Adrien Le Guillou            & Emory University                                    \\
Anna Hotton                  & University of Chicago                               \\
Arindam Fadikar              & University of Chicago/Argonne National Laboratory                          \\
Arvind Ramanathan            & University of Chicago/Argonne National Laboratory                          \\
Aurélien Cavelan             & University of Basel, Switzerland                    \\
Ben Toh                      & Northwestern                                        \\
Benedicta Mensah             & Yale                                                \\
Chick Macal                  & University of Chicago/Argonne National Laboratory                          \\
Clinton Collins              & Bill \& Melinda Gates Foundation                    \\
Jaline Gerardin              & Northwestern                                        \\
James Carson                 & University of Texas at Austin                        \\
Jiangzhuo Chen               & University of Virginia                              \\
Jonathan Ozik                & University of Chicago/Argonne National Laboratory                               \\
Justin Wozniak               & University of Chicago/Argonne National Laboratory                          \\
Kyle Chard                   & University of Chicago/Argonne National Laboratory                             \\
Matteo Turilli               & Rutgers University                                  \\
Maxwell Amon Burnette        & University of Illinois Urbana-Champagne             \\
Mercedes Pascual             & University of Chicago                               \\
Nick Collier                 & University of Chicago/Argonne National Laboratory                          \\
Olya Morozova                & University of Chicago                               \\
Phil Arevalo                 & Wesleyan University                                 \\
Rafael Ferreira da Silva     & Oak Ridge National Laboratory                       \\
Rahul Subramanian            & National Institutes of Health \\
Ravi Madduri                 & University of Chicago/Argonne National Laboratory \\
Rebecca Lee Smith            & University of Illinois Urbana-Champagne             \\
Stefan Hoops                 & University of Virginia                              \\
Tim Germann                  & Los Alamos National Laboratory                      \\
Valerie Hayot-Sasson         & University of Chicago                               \\
Yadu Babuji                  & University of Chicago/Argonne National Laboratory                               \\
Zhaowei Du                   & Bill \& Melinda Gates Foundation                    \\

\bottomrule
\end{tabular}
\end{table}

\newpage
\subsection{Workshop Agenda}\label{sec:agenda}
\begin{table}[h]\label{tbl:agenda}
\centering
\rowcolors{2}{gray!25}{white}
\begin{tabular}{lp{11cm}}
\toprule
\multicolumn{2}{c}{\textbf{Session 1: Evaluate current practices and identify key challenges}} \\
\midrule
\textbf{Time} & \textbf{Activity} \\
\midrule
12:00 - 12:30 & Welcome and lunch \\
12:30 - 13:30 & Panel discussion \newline \textit{Moderated by Jaline Gerardin} \\
13:30 - 13:45 & \textit{Coffee break} \\
13:45 - 16:00 & Breakout groups \newline \textit{Evaluate current workflows, identify key challenges, and envision ideal systems}\\
16:00 - 17:00 & Breakout groups report back \\
17:30 - 20:30 & Dinner \\
\midrule
\multicolumn{2}{c}{\textbf{Session 2: Propose solutions for challenges identified in Session 1}} \\
\midrule
\textbf{Time} & \textbf{Activity} \\
\midrule
08:30 - 09:00 & Breakfast \\
09:00 - 09:15 & Welcome and recap \\
09:15 - 10:30 & Breakout groups \newline \textit{Propose pathways to needed capabilities} \\
10:30 - 11:00 & Breakout group write-ups \\
11:00 - 11:50 & Breakout groups report back \\
11:50 - 12:00 & Wrap-up \\
\bottomrule
\end{tabular}
\label{table:your_label}
\end{table}

\newpage
\subsection{Breakout Session 1 Questions}\label{breakout1_qs}

\subsubsection*{\textit{Evaluation phase} (1 hr)} 

\begin{enumerate}
\item Evaluating current computational resources
\begin{itemize}
\itemsep0em 
\item What computational resources do you use?  
\item Where are they located?  
\item What type (e.g., HPC, cloud)?  
\item What attempts, if any, have you made to scale your code or streamline your analyses? For example, 
\begin{itemize}
\item Have you used any libraries or workflow systems? 
\item Do you use GPUs or other specialized hardware? 
\end{itemize}
\item What are the strengths and limitations of your current approaches and computing infrastructure? What would you most want to change about your current workflow or computing infrastructure? What are the questions you currently can’t answer that you’d like to? 
\end{itemize}

\item Evaluating current data practices 
\begin{itemize}
\itemsep0em 
\item What data do you use?  
\item Where is the data located?  
\item How is it accessed?   
\item What methods and technologies do you use to collate and harmonize data (i.e., across the different temporal and spatial scales of monitoring environmental, animal, and human health and contextual data)? 
\item What are the strengths and limitations of your current approaches to accessing, retrieving, and using data? What would you most want to change about your current approach to data? 
\end{itemize}

\item Evaluating technical support 
\begin{itemize}
\item Do you currently have access to HPC experts to answer questions and help with implementation? 
\end{itemize}
\end{enumerate}

\paragraph{\textit{Visioning phase} (45 mins)}  

\begin{itemize}
\item In an ideal world, what would you want to be able to do, modeling-wise?  
\item In an ideal world, how would you like to store and access data?  
\item In an ideal world, what kind of technical support do you have access to? 
\end{itemize}

\paragraph{\textit{Drawing phase} (30 mins) }
How would you describe your ideal end-to-end compute systems?  

\begin{itemize}
\item Describe its purpose (e.g., forecasting, scenario running) 
\item Describe its user(s) 
\item Use 3 words to describe its most important performance features 
\item Draw it: what should the components be, how should they be connected, what should the inputs and outputs be, what should key features of each component or connection be (technical or performance-based descriptors, adjectives like “fast” or “reliable” ok too), etc 
\end{itemize}

\end{appendices}
\end{document}